\documentclass[a4paper]{ISOStyle}
\usepackage{graphicx}

\begin{document}

\title{Radiative pumping of 1612\,MHz OH masers: OH/IR sources
with IRAS LRS spectra and 34.6 micrometer absorption feature.
}

\author{Ryszard Szczerba\inst{1}, Jin Hua He\inst{2} 
\and Pei Sheng Chen\inst{2}} 
  
\institute{N. Copernicus Astronomical Center, Rabia{\'n}ska 8, 
87-100 Toru{\'n}, Poland
  \and 
  Yunnan Astronomical Observatory, National Astronomical Observatories, CAS, 
Kunming 650011, P.R. China
}

\maketitle 

\begin{abstract}
The population inversion which leads to the 1612 MHz OH maser emission has 
long been thought to be radiatively pumped. Since OH rotational lines 
involved in this pumping scheme lie in the far--infrared they became 
observable only after the launch of the ISO satellite. With the aim to 
investigate the 
pumping conditions of the 1612 MHz OH maser emission in more details we have 
searched the ISO Archive for SWS observations around 34.6\,$\mu$m of
1024 OH/IR sources with IRAS LRS spectra from compilation of Chen et 
al.\,(2001). Surprisingly, among 81 OH/IR sources which have appropriate
SWS data only already reported objects: VY\,CMa, IRC$+$10420 and the 
Galactic center, show clear 34.6\,$\mu$m absorption line. 
We discuss possible reasons for non--detection of this pumping line.

\end{abstract}

\section{INTRODUCTION}

The first detection of intense radio emission from OH molecules was reported 
by Weaver et al.\,(1965) and soon an explanation based on maser amplification 
through induced processes was invoked (Litvak et al.\,1966, Perkins et 
al.\,1966). Shklovsky\,(1966) was first who proposed a radiative pumping 
mechanism for OH masers and his idea was elaborated in detail by Elitzur et 
al.\,(1976) for infrared (IR) stars which exhibit the 1612 MHz OH satellite 
line (hereafter OH/IR stars). In this scheme the required inversion 
of $F$=1 and $F$=2 sub-levels (even and odd parity, respectively) in the 
lowest rotational level is achieved by absorption of infrared photons 
(predominantly at 34.6\,$\mu$m) from the ground state of OH molecule
($^2{\Pi}_{3/2}\,J$=3/2) and consequent radiative decays to lower levels
via other far-infrared (FIR) transitions.

Due to atmospheric absorption in the FIR region, OH rotational lines are
inaccessible from the ground and only indirect checks of the pumping
scheme  was possible using infrared flux extrapolated to about 35\,$\mu$m.
For example, it was shown that there are enough FIR photons (about 4 FIR 
photons for one OH photon - Elitzur 1992) to pump
the 1612 MHz OH maser (Evans \& Beckwith\,1977, Nguyen-Q-Rieu et al.\,1979,
Epchtein et al.\,1980). Direct confirmation of this theory become possible 
only 
with the launch of the Infrared Space Observatory (ISO, Kessler et al.\,1996) 
which allows observation of the pumping transition(s) and other
FIR OH lines from OH maser sources. First clear absorption at 34.6\,$\mu$m in
the Short Wavelength Spectrometer (SWS, de Graauw et al. 1996) spectrum
of supergiant NML~Cyg was reported by Justtanot et al. (1996) and in case of
the Galactic center by Lutz et al. (1996). A detailed analysis supporting 
the radiative pumping cycle for 
circumstellar masers have been presented by Sylvester et al.\,(1997) in the
case of another supergiant IRC$+$10420. They detected not only the pumping line
at 34.6\,$\mu$m in the SWS\,02 spectrum of IRC$+$10420 (now resolved into 
$\Lambda$-doublet components),
but also rotational 
cascade lines at 98.7, 163.1 and 79.2\,$\mu$m. Thai-Q-Tung et al.\,(1998) 
performed modeling of pumping conditions and maser radiative 
transfer calculations for this supergiant. Their results are in 
agreement with the observed FIR OH lines, confirming 
the theoretical pump scheme. There are three more sources for
which published SWS observations show absorption of the 1612 MHz pumping line
at 34.6\,$\mu$m. Namely, Neufeld et al.\,(1999) showed that this line is seen 
in the supergiant star VY CMa, Skinner et al.\,(1997) have reported its 
detection towards the ultra-luminous infrared galaxy Arp\,220 and Bradford et
al.\,(1999) have observed it in the starbust galaxy NGC\,253.
Summarizing, up to now there are 6 published detections of 34.6\,$\mu$m 
absorption line from OH maser sources (from 3 supergiants: \object{NML~Cyg}, 
\object{IRC$+$10420}, \object{VY~CMa}, from 2 galaxies: \object{Arp~220},
\object{NGC~253} and from the Galactic center).

In this paper we report on a search of the ISO Data Archive (IDA) for further 
evidence of absorption at 34.6 $\mu$m among OH/IR sources which were
observed with the ISO SWS. In Section\,2 we describe our working sample, 
present the 1612 MHz OH/IR sources which have available SWS observations 
around 34.6\,$\mu$m, discuss briefly the SWS data reduction process and
present examples of the SWS spectra for OH/IR sources with detected 
absorption line at 34.6\,$\mu$m. Finally, in Section\,3 we discuss 
briefly the results obtained.

\section{OBSERVATIONS}
%
%
\begin{table}[]{}
   \caption[]{Results of the IDA search for SWS data covering 34.6\,$\mu$m 
region within 1 arcmin around IRAS position of OH/IR sources from list of 
Chen et al.\,(2001).}
  \begin{center}
  \leavevmode
  \footnotesize
    \begin{tabular}{l l l l l l l}\hline
            \noalign{\smallskip}
IRAS         & LRS   & IDA            & \multicolumn{4}{c}{SWS~AOT} \\
name         & class & name           & 01  & 02  & 06  & 07 \\
            \noalign{\smallskip}
            \hline
            \noalign{\smallskip}
\object{01037$+$1219} & E & WX Psc             & +   & +   & +   & +   \\
\object{01304$+$6211} & A & OH 127.8+0.0       & +   & $-$ & $-$ & $-$ \\
\object{02192$+$5821} & E & S Per              & +   & $-$ & +   & $-$ \\
\object{03507$+$1115} & E & IK Tau             & $-$ & $-$ & +   & +   \\
\object{05073$+$5248} & E & IRC+50137          & +   & $-$ & $-$ & $-$ \\
\object{05373$-$0810} & C &                    & +   & $-$ & $-$ & $-$ \\
\object{05380$-$0728} & U &                    & +   & $-$ & $-$ & $-$ \\
\object{05506$+$2414} & H &                    & +   & $-$ & $-$ & $-$ \\
\object{06053$-$0622} & H & MON R2 IRS3        & +   & $-$ & +   & $-$ \\
\object{06238$+$0904} & C &                    & +   & $-$ & $-$ & $-$ \\
\object{07027$-$7934} & H &                    & +   & $-$ & $-$ & $-$ \\
\object{07209$-$2540} & E & VY CMa             & +   & $-$ & +   & +   \\
\object{10197$-$5750} & H & Roberts 22         & +   & $-$ & $-$ & $-$ \\
\object{10580$-$1803} & E & R Crt              & $-$ & +   & $-$ & $-$ \\
\object{13517$-$6515} & E &                    & +   & $-$ & $-$ & $-$ \\
\object{15452$-$5459} & U &                    & +   & $-$ & $-$ & $-$ \\
\object{15559$-$5546} & P &                    & +   & $-$ & $-$ & $-$ \\
\object{16235$+$1900} & E & BS 6119            & +   & $-$ & $-$ & $-$ \\
\object{16279$-$4757} & H &                    & +   & $-$ & $-$ & $-$ \\
\object{16280$-$4008} & H & NGC 6153           & +   & +   & $-$ & $-$ \\
\object{16342$-$3814} & H &                    & +   & $-$ & $-$ & $-$ \\
\object{17004$-$4119} & A &                    & +   & $-$ & $-$ & $-$ \\
\object{17010$-$3840} & A &                    & +   & $-$ & $-$ & $-$ \\
\object{17103$-$3702} & L & NGC 6302           & +   & +   & +   & $-$ \\
\object{17150$-$3224} & H &                    & +   & $-$ & $-$ & $-$ \\
\object{17319$-$6234} & E &                    & +   & $-$ & $-$ & $-$ \\
\object{17347$-$3139} & H &                    & +   & $-$ & $-$ & $-$ \\
\object{17360$-$3012} & E &                    & +   & $-$ & $-$ & $-$ \\
\object{17393$-$3004} & U & 1742$-$3005        & +   & $-$ & $-$ & $-$ \\
\object{17411$-$3154} & A & AFGL 5379          & +   & $-$ & $-$ & $-$ \\
\object{17418$-$2713} & A &                    & +   & $-$ & $-$ & $-$ \\
\object{17424$-$2859} & H & GC Sgr\,A$^*$      & +   & +   & +   & +   \\
\object{17430$-$2848} & H & GCS 3 I            & +   & $-$ & $-$ & $-$ \\
\object{17431$-$2846} & H & G0.18$-$0.04       & +   & $-$ & $-$ & $-$ \\
\object{17443$-$2949} & A &                    & +   & $-$ & $-$ & $-$ \\
\object{17463$-$3700} & F & H1$-$36            & +   & $-$ & $-$ & $-$ \\
\object{17501$-$2656} & E & AFGL 2019          & +   & $-$ & $-$ & $-$ \\
\object{17516$-$2526} & U &                    & +   & $-$ & $-$ & $-$ \\
\object{17554$+$2946} & E & AU Her             & +   & $-$ & $-$ & $-$ \\
\object{17574$-$2403} & H & W\,28A2            & +   & $-$ & $-$ & $-$ \\
\object{18050$-$2213} & E & VX Sgr             & +   & $-$ & $-$ & $-$ \\
\object{18095$+$2704} & H &                    & +   & $-$ & $-$ & $-$ \\
\object{18123$+$0511} & F &                    & +   & $-$ & $-$ & $-$ \\
\object{18139$-$1816} & I & OH 12.8$-$0.9      & +   & $-$ & $-$ & $-$ \\
\object{18196$-$1331} & A & GL 2136            & +   & $-$ & +   & $-$ \\
\object{18257$-$1000} & A & OH 21.5+0.5        & +   & $-$ & $-$ & $-$ \\
\object{18276$-$1431} & H & OH 17.7$-$2.0      & +   & $-$ & $-$ & $-$ \\
\object{18348$-$0526} & A & OH 26.5+0.6        & +   & +   & $-$ & $-$ \\
\object{18349$+$1023} & E & V1111 Oph          & $-$ & +   & +   & $-$ \\
\object{18385$-$0617} & A & OH 26.2$-$0.6      & +   & $-$ & $-$ & $-$ \\
\object{18460$-$0254} & A & OH 30.1$-$0.7      & +   & $-$ & $-$ & $-$ \\
\object{18488$-$0107} & A & OH 32.0$-$0.5      & +   & $-$ & $-$ & $-$ \\
\object{18498$-$0017} & H & OH 32.8$-$0.3      & +   & $-$ & $-$ & $-$ \\
\object{18549$+$0208} & A & OH 35.6$-$0.3      & +   & $-$ & $-$ & $-$ \\
            \noalign{\smallskip}
            \hline
         \end{tabular}
  \end{center}
\end{table}
%
%
\setcounter{table}{0}
\begin{table}[]{}
   \caption[]{continuation.}
  \begin{center}
  \leavevmode
  \footnotesize
    \begin{tabular}{l c l c c c c}\hline
            \noalign{\smallskip}
IRAS         & LRS   & IDA            & \multicolumn{4}{c}{SWS~AOT} \\
name         & class & name           & 01  & 02  & 06  & 07 \\
            \noalign{\smallskip}
            \hline
            \noalign{\smallskip}
\object{18551$+$0323} & C &                    & +   & $-$ & $-$ & $-$ \\
\object{18556$+$0811} & E & EIC 722            & +   & $-$ & $-$ & $-$ \\
\object{18560$+$0638} & A & OH 39.7+1.5        & +   & $-$ & $-$ & $-$ \\
\object{19039$+$0809} & E & R$-$Aql            & +   & $-$ & +   & $-$ \\
\object{19114$+$0002} & H & HD 179821          & +   & +   & $-$ & $-$ \\
\object{19192$+$0922} & A & OH 44.8$-$2.3      & +   & $-$ & $-$ & $-$ \\
\object{19219$+$0947} & H & VY 2$-$2           & +   & +   & $-$ & $-$ \\
\object{19244$+$1115} & E & IRC+10420          & +   & +   & +   & +   \\
\object{19255$+$2123} & H & PK 056+2.1         & +   & $-$ & $-$ & $-$ \\
\object{19283$+$1944} & A &                    & +   & $-$ & $-$ & $-$ \\
\object{19327$+$3024} & P & BD +30 3639        & +   & +   & $-$ & $-$ \\
\object{19343$+$2926} & H & M1$-$92            & +   & +   & $-$ & $-$ \\
\object{19550$-$0201} & E & RR Aql             & +   & $-$ & $-$ & $-$ \\
\object{20000$+$4954} & E & Z Cyg              & +   & $-$ & $-$ & $-$ \\
\object{20043$+$2653} & A &                    & +   & $-$ & $-$ & $-$ \\
\object{20077$-$0625} & E & IRC$-$10529        & +   & $-$ & $-$ & $-$ \\
\object{20255$+$3712} & H & S 106              & +   & +   & $-$ & $-$ \\
\object{20272$+$3535} & A &                    & +   & $-$ & $-$ & $-$ \\
\object{20406$+$2953} & H &                    & +   & $-$ & $-$ & $-$ \\
\object{20529$+$3013} & E & UX Cyg             & +   & $-$ & $-$ & $-$ \\
\object{21554$+$6204} & A &                    & +   & $-$ & $-$ & $-$ \\
\object{22036$+$5306} & U & HD 235718          & +   & $-$ & $-$ & $-$ \\
\object{22176$+$6303} & H & S 140              & +   & +   & $-$ & +   \\
\object{22177$+$5936} & A & OH 104.9+2.4       & +   & $-$ & $-$ & $-$ \\
\object{22556$+$5833} & E & AFGL 2999          & +   & $-$ & $-$ & $-$ \\
\object{23412$-$1533} & E & R Aqr              & +   & $-$ & $-$ & $-$ \\
\object{23416$+$6130} & E & PZ Cas             & +   & $-$ & $-$ & $-$ \\
            \noalign{\smallskip}
            \hline
         \end{tabular}
  \end{center}
\end{table}
Chen et al.\,(2001) discussed properties of the 1612 MHz OH sources associated
with the InfraRed Astronomical Satellite (IRAS) counterparts which have Low
Resolution Spectra (LRS) available. Altogether this sample consists of 1024 
OH/IR sources for which the difference between OH maser and IRAS position is 
smaller than 1\arcmin. These sources were classified according to the Volk \& 
Cohen\,(1989) classification scheme and it was found that sources with 
silicate emission (class E) form the largest group (about 57\%) followed
by the group with silicate absorption (class A: about 16\%) and by group of 
sources with red-continuum (class H: about 6\%). Information about LRS 
classification (i.e. about optical depth at least in the case of E and A 
sources) 
was intended to be used for interpretation of the 34.6\,$\mu$m absorption line 
detection frequency in our sample of OH/IR sources. 

We have searched the IDA for SWS data within 1 arc-min around IRAS position 
of 1024 galactic OH/IR sources from Chen et al.\,(2001) list (NML Cyg is not 
included in our sample as there is no IRAS observations for this supergiant,
while Arp\,220 and NGC\,253 are extragalactic mega-maser sources). 
The results of our 
search for SWS data around 34.6\,$\mu$m are given in Table 1. 
The associated IRAS name is given in column 1, IRAS LRS spectrum classification
in column 2 and most frequently used source name from the 
original ISO proposals (if different from the IRAS one) in column 3. Sign 
$+$ in columns 4, 5, 6 and 7 means that at least one SWS spectrum, covering 
wavelengths range around 34.6\,$\mu$m, taken with Astronomical Observation 
Template (AOT) 01, 02, 06 and 07, respectively, is available. In some cases
(e.g. GC Sgr\,A$^*$) there are many SWS spectra available inside a 1\arcmin\
circle around IRAS position. All of them were carefully checked but a complete 
analysis will be published elsewhere.

The ISO SWS\,01, 02, 06 and 07 data (offline processing -- OLP version 9.5)
analyzed in this work were all processed using ISAP (ISO Spectroscopic 
Analysis Package) version 2.1. Recently, a new versions of OLP (10.0, 10.1) 
have been released, but these newer data should not change our conclusions as 
far as the detection/non-detection frequency of the 34.6\,$\mu$m absorption 
line is concerned. Data analysis
consisted of extensive bad data removal primarily to minimize the effects of 
cosmic rays. First, all detectors were compared to identify possibly narrow 
features and then the best detector was chosen to compare one by one with 
others during the process of bad data removal. However, there are 
usually fewer scans in SWS\,07 data than in other AOT's, so all scans were 
processed simultaneously in these cases. In the next step the spectra were 
scaled to the same flux level to correct for different responsivities of the 
detectors and 
any remaining outliers removed. When scaling, spectra were shifted by the 
{\it offset} mode if their overall flux density was lower than about 
100\,Jy, while 
by the {\it gain} mode when they had flux density higher than this limit. 
For SWS\,06 spectra, the two scan directions correspond to two different lines
and it is not possible to shift them simultaneously within ISAP. Therefore, 
scaling was done for the two directions separately. In addition, 
whenever memory 
effects or irregularities were present in the two scans of SWS\,01 or 02 data,
we averaged them separately and the resulting two sub--spectra were 
used to check reality of possible features. Finally, spectra were averaged 
across detectors, scans and lines (if applicable), using median clipping to 
discard points that lay more than 2.5$\sigma$ from the median flux. 
Spectra were averaged typically to 
resolution of 400, 500, 800 and 1500 for SWS\,01 data taken with speed 01, 02,
03 and 04, respectively, and to resolution of 3000 (SWS\,02), 1500 (SWS\,06) 
and 30000 (SWS\,07) for the other observation modes. Finally, when the 
34.6\,$\mu$m absorption line was 
detected we used ISAP to determine continuum level and fit a single Gaussian 
profile to estimate line parameters. Only in case of the 34.6\,$\mu$m line 
which was resolved into $\Lambda$--doublet components we fit Gaussian profile 
to each of them.

\begin{figure}[]
\resizebox{\hsize}{!}{\includegraphics[angle=90]{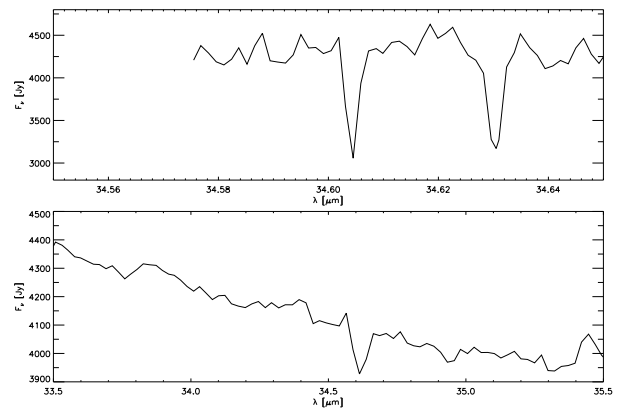}}
\caption{SWS\,07 (upper panel) and SWS\,06 (bottom panel) spectrum of 
supergiant VY CMa (IRAS 07209$-$2540) around 34.6\,$\mu$m.}
\end{figure}
\begin{figure}[]
\resizebox{\hsize}{!}{\includegraphics[angle=90]{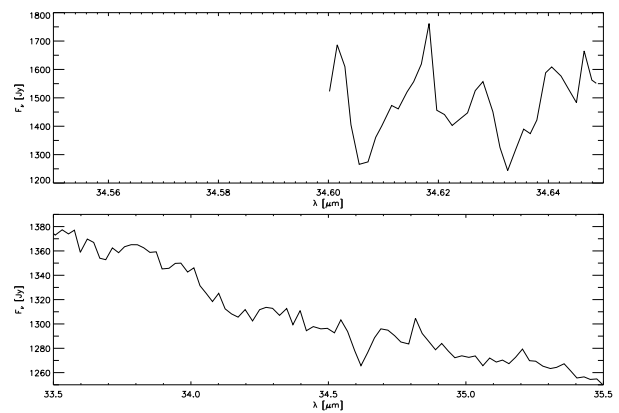}}
\caption{SWS\,07 (upper panel) and SWS\,06 (bottom panel) spectrum of 
supergiant IRC$+$10420 (IRAS 19244$+$1115) around 34.6\,$\mu$m.}
\end{figure}
\begin{figure}[]
\resizebox{\hsize}{!}{\includegraphics[angle=90]{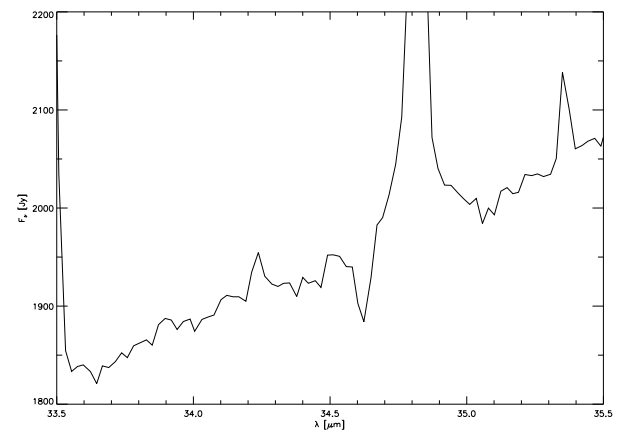}}
\caption{SWS\,01 spectrum of Sgr A$^*$ (IRAS 17424$-$2859) 
around 34.6\,$\mu$m.}
\end{figure}

Altogether, we have processed 159 SWS spectra (114 for AOT 01; 21 for 02; 
16 for 06 and 8 for 07) around 34.6\,$\mu$m for 
81 OH/IR sources which have IRAS LRS spectra. The 34.6\,$\mu$m absorption 
feature was undoubtedly detected in the SWS spectra taken toward {\it only} 
3 OH/IR sources from our sample: supergiant VY CMa (IRAS 07209
$-$2540); supergiant IRC$+$10420 
(IRAS 19244$+$1115) and the Galactic center (IRAS 17424$-$2859). In Figs.1--3
we present examples of unpublished spectra for these 3 sources. 
A detailed discussion of all detections (including tentative ones) for 
the more complete sample of OH/IR sources will 
appear elsewhere. Information about presented spectra, the obtained number of 
34.6\,$\mu$m photons (n$_{34.6}$) and number of the OH photons (n$_{OH}$) are 
given in Table\,2.

\begin{table}[]
  \caption{Observational details for sources with detected 
34.6\,$\mu$m absorption line.
}
  \begin{center}
  \leavevmode
  \footnotesize
  \begin{tabular}{l l l l l l}\hline 
            \noalign{\smallskip}
source  & Obs. & TDT    & n$_{34.6}$\,10$^5$ &  n$_{OH}$\,10$^2$ \\
name    & mode     &        &  [m$^{-2}$s$^{-1}$] & [m$^{-2}$s$^{-1}$] \\
         \noalign{\smallskip}
         \hline
         \noalign{\smallskip}
VY CMa      & 07     & 73601963  & 14.1+17.9 & 1597$^1$ \\
            & 06     & 73402218  & 45.8 &      \\
IRC$+$10420 & 07     & 36401613  & 10.4+9.9 & 620$^2$ \\
            & 06     & 31600936  & 7.3 &    \\
GC Sgr A$^*$   & 01(04) & 09500203  & 18.4 & 41.1$^3$ \\
         \noalign{\smallskip}
         \hline
         \noalign{\smallskip}
     \end{tabular}
\begin{list}{}{}
\item[]
$^1$Neufeld et al. (1999); $^2$Sylvester et al.\,(1997); $^3$Derived from 17
maser sources (Sjouwerman et al.\,1998) which are located within 1\arcmin 
around IRAS\,17424$-$2859 position.
For SWS\,07 observations  n$_{34.6}$ is given as a sum of two numbers
which correspond to each $\Lambda$--doublet component. 
Number in parenthesis denote speed 
of SWS\,01 observations.
\end{list}  
\end{center}
\end{table}

\section{DISCUSSION}

As we discussed in Introduction the theory of radiative pumping of the 1612 
MHz OH maser emission seems to be well established. In addition, 
pump rates (n$_{OH}$/n$_{34.6}$) determined from the ISO SWS spectra 
analyzed here 
(5\% -- TDT 73601963 and 3.5\% -- TDT 73402218  for VY CMa; 
3.1\% -- TDT 36401613 and 8.5\% -- TDT 31600936 for IRC$+$10420 but only
0.2\% -- TDT 09500203 for the Galactic center) do not contradict the proposed 
pumping scheme which requires about 4 FIR photons for one OH photon. 
Therefore, it is surprising that there are {\it only} 3 cases among 81 OH/IR 
sources with appropriate ISO SWS data which show a clear absorption at 
34.6\,$\mu$m.

Certainly, the detection rate of the 34.6\,$\mu$m pumping line depends on the
signal to noise ratio of the SWS spectra. Sources with tentative detection
of this absorption line (not discussed here) have rather noisy spectra and it 
is difficult to prove the reality of the line. In any case, the number of 
sources with
tentative detections is rather small and low signal to noise ratio does not 
probably solve the problem of the 34.6\,$\mu$m line non--detection in other 
sources with good signal-to-noise SWS spectra around 35\,$\mu$m.  
Another factor which could 
influence the detection of this absorption feature is the spectral resolution. 
However, among analyzed data there are spectra with high resolution 
(SWS\,01 speed 4, 
SWS\,02, SWS\,06 and, especially, SWS\,07) which do not show absorption at 
34.6\,$\mu$m. The detection rate also does not depend on the source flux. In 
our 
sample we have sources with flux at 34.6\,$\mu$m well in excess of 1000 Jy 
and still there is no signature of {\it any} absorption at 34.6\,$\mu$m. 
Therefore, we believe that explanation of this puzzling result is related to 
the geometry (relative location and size) of masering and dusty regions and/or
to differences in physical conditions inside circumstellar (or interstellar) 
shells (clouds). Possibly, regions containing masering OH molecules (OH spots)
are much smaller than regions emitting IR photons and absorption at 
34.6\,$\mu$m is filled by more spatially extended IR emission.
However, the fact that the 34.6\,$\mu$m absorption 
line is seen {\it only} in supergiants and in the extragalactic sources 
(Galactic center is very likely to be similar to the later cases) 
is probably not a 
coincidence and requires more careful analysis of physical conditions around
sources with detected and non--detected 34.6\,$\mu$m pumping line. 
More detailed discussion based on sample of all known galactic OH/IR sources 
with IRAS identification will appear elsewhere.

\begin{acknowledgements}

This work has been partly supported by grant 2.P03D.024.18p01 of the Polish 
State Committee for Scientific Research and by the Chinese Academy of 
Sciences and the NNSF of China.

\end{acknowledgements}

\end{document}